\documentclass{article}

\usepackage{arxiv}

\usepackage[utf8]{inputenc} 
\usepackage[T1]{fontenc}    
\usepackage{hyperref}       
\usepackage{url}            
\usepackage{booktabs}       
\usepackage{amsfonts}       
\usepackage{nicefrac}       
\usepackage{microtype}      
\usepackage{lipsum}
\usepackage{graphicx}
\usepackage{subfigure}
\usepackage{amsmath}
\graphicspath{ {./images/} }

\title{Generative and Discriminative Learning for Distorted Image Restoration}

\author{
	Yi Gu$^{\dagger}$ \qquad Yuting Gao$^{\ddag}$ \qquad Jie Li$^{\dagger}$ $^{\ast}$ \qquad Chentao Wu$^{\dagger}$ \qquad  Weijia Jia$^{\S}$\\
 $^{\dagger}$ Department of Computer Science and Engineering, Shanghai Jiao Tong University, Shanghai, China\\
$^{\ddag}$ Department of Electrical and Computer Engineering, Texas A\&M University, College Station, US\\
$^{\S}$ State Key Laboratory of Internet of Things for SmartCity, University of Macau, Macau, China\\
\texttt{\{louisgu,lijiecs\}@sjtu.edu.cn}
}

\begin{document}
\maketitle
\begin{abstract}
Liquify is a common technique for image editing, which can be used for image distortion. Due to the uncertainty in the distortion variation, restoring distorted images caused by liquify filter is a challenging task. To edit images in an efficient way, distorted images are expected to be restored automatically. This paper aims at the distorted image restoration, which is characterized by seeking the appropriate warping and completion of a distorted image. Existing methods focus on the hardware assistance or the geometric principle to solve the specific regular deformation caused by natural phenomena, but they cannot handle the irregularity and uncertainty of artificial distortion in this task. To address this issue, we propose a novel generative and discriminative learning method based on deep neural networks, which can learn various reconstruction mappings and represent complex and high-dimensional data. This method decomposes the task into a rectification stage and a refinement stage. The first stage generative network predicts the mapping from the distorted images to the rectified ones. The second stage generative network then further optimizes the perceptual quality. Since there is no available dataset or benchmark to explore this task, we create a Distorted Face Dataset (DFD) by forward distortion mapping based on CelebA dataset. Extensive experimental evaluation on the proposed benchmark and the application demonstrates that our method is an effective way for distorted image restoration.
\end{abstract}


\begin{figure}[h]
	\centering
	\subfigure[]{
		\includegraphics[width=2.5cm,height=2cm]{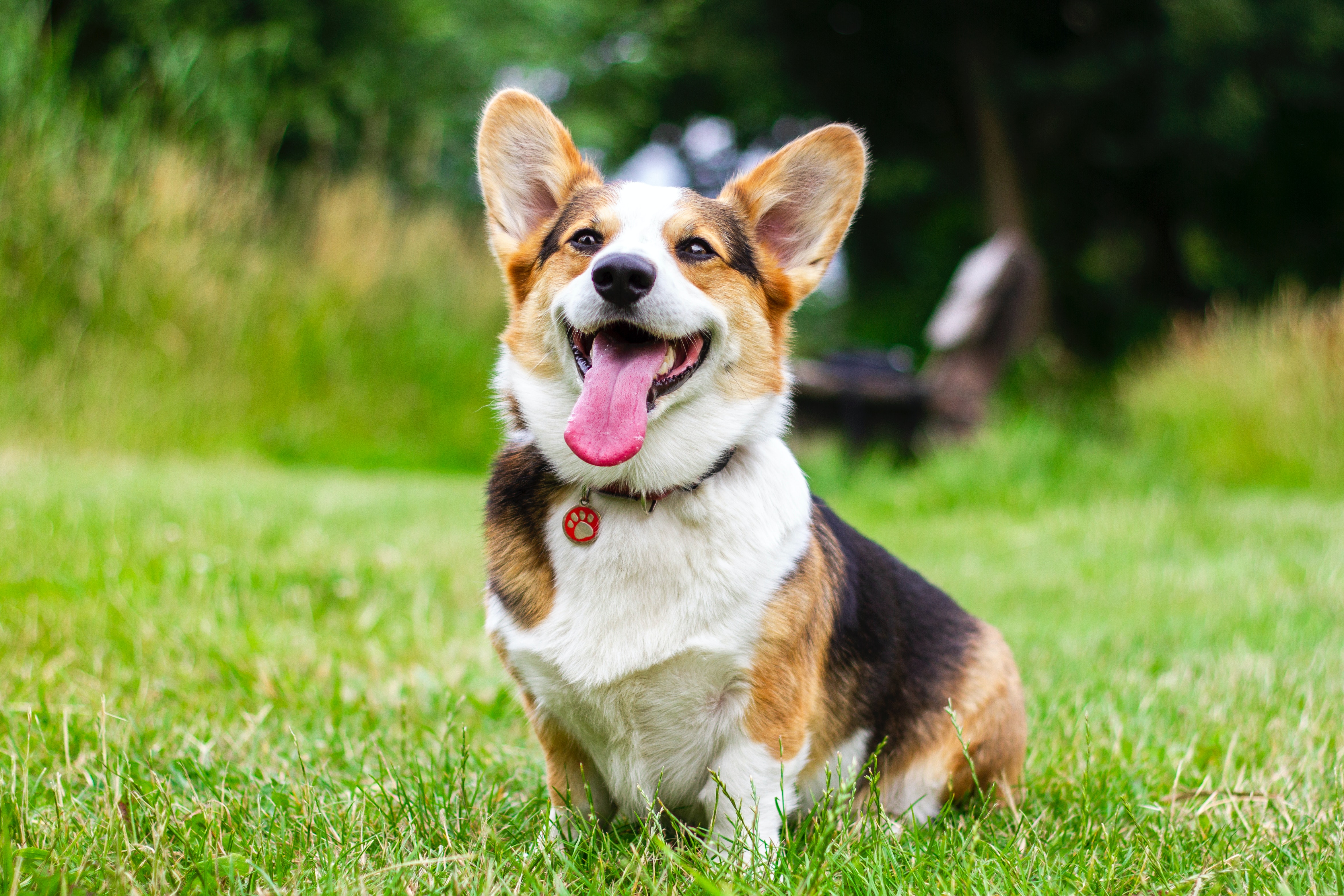}
	}
	\subfigure[]{
		\includegraphics[width=2.5cm,height=2cm]{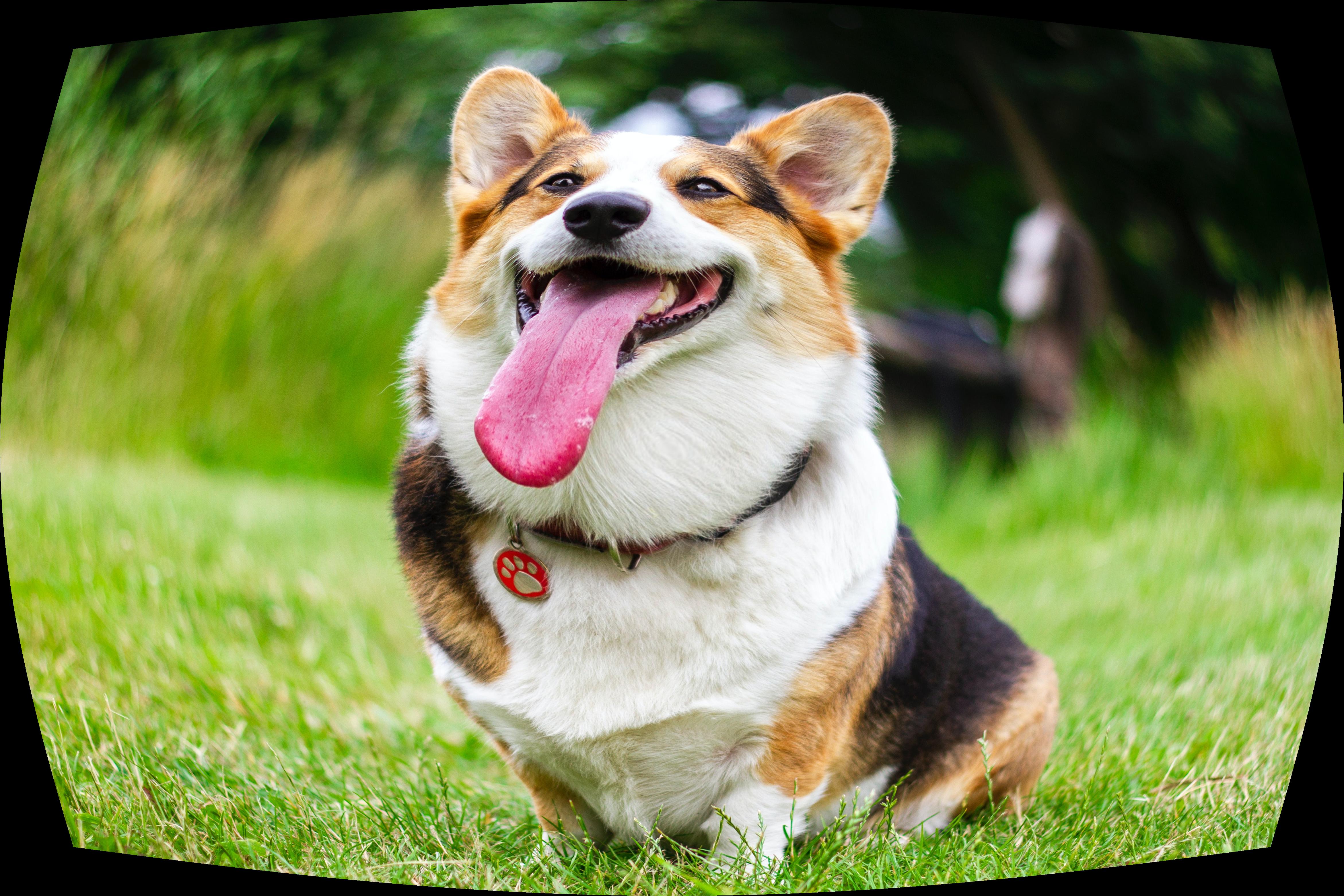}
	}
	\subfigure[]{
		\includegraphics[width=2.5cm,height=2cm]{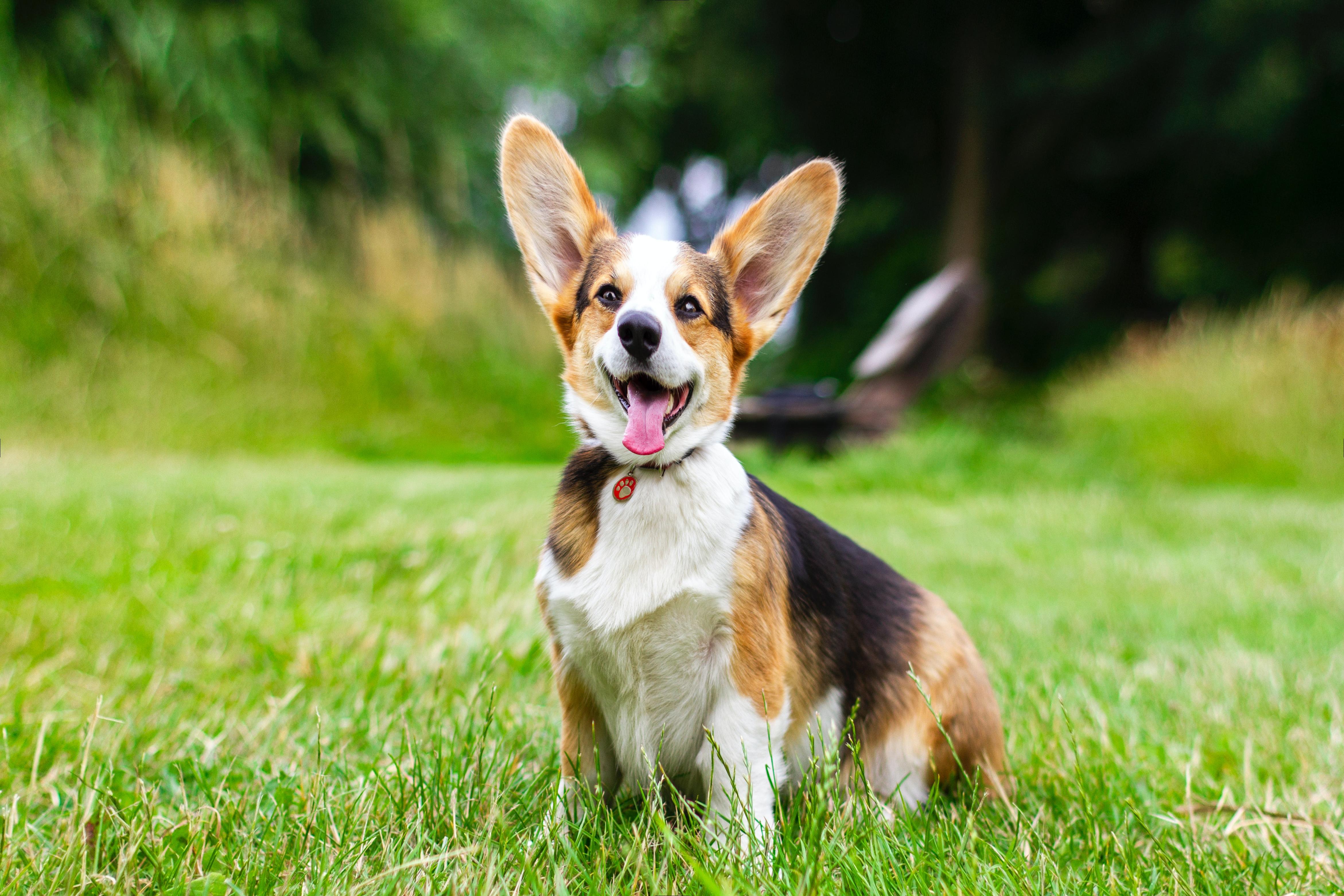}
	}
	\caption{For a given original image (a), we show the convex lens effect (b) and the concave lens effect (c). The distorted image restoration task is to seek the appropriate warping of the distorted image and restore it to the original image.}	
	\label{task}
\end{figure}

\begin{figure*}
	\centering
	\includegraphics[scale=0.7]{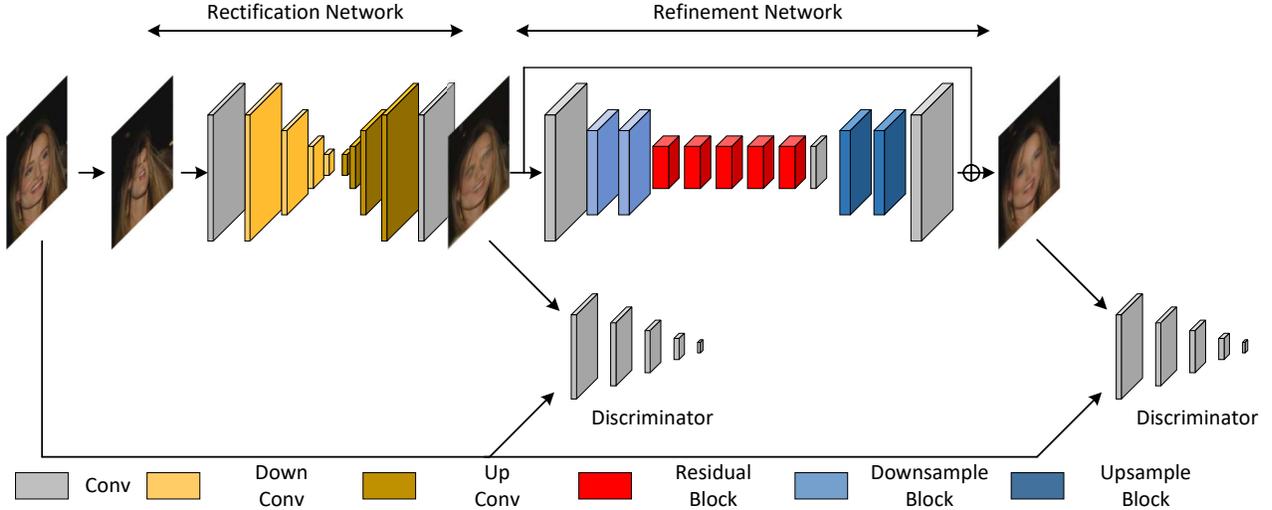}
	\caption{Framework Overview. Given a distorted image, the simply corrected image is first generated by rectification network, then the refinement network further produces the missing content to make the prediction more realistic and coherent. Two discriminators learn to distinguish the generated image from the original image.}
	\label{framework}
\end{figure*}

\section{Introduction}
Liquify \cite{cromhout2003liquifying} is significantly valued to the applications in image editing. As shown in Figure \ref{task}, the liquify filter can be used to distort the image to achieve various special effects, such as concave lens, convex lens, distorting mirror, etc. Such image effects are widely used in art and entertainment applications. Although we can use the image editing application to liquify the image, it is difficult to restore the distorted image. When the image editor desires to remove special effects from distorted image material, it often takes them much time to restore the material by liquify operation. To this end, we expect an automatic distorted image restoration method. Users only need to upload a distorted image, and they will get the rough original image with the distorted special effect removed. This solution can reduce the workload of image editors and facilitate them to do more refined operations on images.

In this paper, the distorted image restoration task we discussed is a relative novel work. Previous efforts mainly focus on automatically restoring the regular deformation, which occurs in natural scenes such as photographs and scans. One common practice \cite{hirano20143d,meng2014active,brown2001document,zhang2008improved} is to rely on additional information acquired by the camera or the sensor to reconstruct the 3D distortion shape. However, it requires complex hardware settings and cannot be applied on a large variation scale, which limits the practical application. Some others \cite{meng2015extraction, meng2018exploiting, moravec2016automatic,yue2019distorted, zhang2019line} correct images based on geometric algorithms. These algorithms are designed for specific types of deformation respectively, but this task requires a method to process multiple types of distortion simultaneously. Compared with the traditional geometric deformation problems, the challenges of this task can be summarized as two aspects: 1) The degrees of distortion in image editing are irregular and unnatural. 2) The types of distortion are various and unpredictable. Therefore, it is more difficult to remove the liquify effect in image editing, which requires the solution to handle the irregularity and complexity of the distortion in this task.

Motivated by the success of applying generative adversarial networks (GANs) \cite{goodfellow2014generative} in image generation \cite{heim2019constrained}, image super-resolution \cite{jiang2019edge}, style transfer \cite{lin2020gan} and representation learning \cite{mathieu2016disentangling}, we present a novel restoration method by generative and discriminative learning to alleviate the above issues. We utilize the generator to reconstruct the mapping from the distorted images to the original ones and produce high frequency details. Besides, the discriminator is adopted to ensure that the generated image is perceptually convincing. This data-driven method can better generalize on various types of distorted image if there is enough training data.

We formulate this task as finding the appropriate warping and completion of a distorted image. To achieve this goal, we define the solution as a two-stage process which involves a rectification stage and a refinement stage. The first stage is designed to rectify the image to an undistorted state. We train a rectification network to rough out the corrected image. The second stage is proposed to supplement image details. A refinement network guides this stage to complete the missing content in rough predictions. We regularize the training process of each stage by defining the loss as the summation of the content loss and the adversarial loss, which ensures the image to render more realistic.

Since there is no existing benchmark dataset available to evaluate distorted image restoration, we create a Distorted Face Dataset (DFD) based on the CelebA \cite{liu2015faceattributes} dataset. The dataset contains 202,599 images with large variations in degree and type of distortion. It can be used as a benchmark to fill the vacuum. The contributions in this paper include:
\begin{itemize}
	\item We perform an initial study of generative and discriminative learning performance on distorted image restoration. We define a generative restoration model that consists of a rectification stage and a refinement stage. Our model establishes a baseline for the task and gives us a view of the current achievements and bottlenecks.
	
	\item We release a Distorted Face Dataset (DFD) as a benchmark for distorted image restoration task. To the best of knowledge, it is the first publicly available dataset on this domain.
\end{itemize}

\begin{figure*}
	\centering
	\includegraphics[scale=0.6]{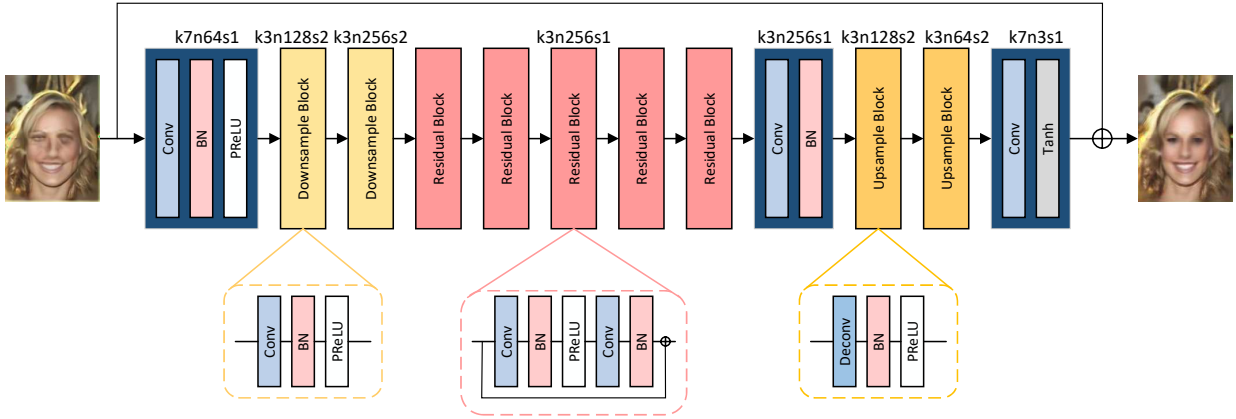}
	\caption{An overview of the refinement network with corresponding kernel size (k), channel number (n) and stride (s) indicated for each block. It contains two downsample blocks, five residual blocks \cite{he2016deep} and two upsample blocks.}
	\label{refinement}
\end{figure*}
\section{Related Work}
Traditional approaches for distorted image restoration include the hardware assistance and the geometric principle. The former relies on the hardware to attain additional information, such as BFS-Auto \cite{hirano20143d}, laser beams \cite{meng2014active}, cameras \cite{brown2001document}, depth sensors \cite{zhang2008improved}, etc. These efforts use the device to reconstruct 3D shapes. Ulges et al. \cite{ulges2004document} managed to adopt general purpose stereo vision methods without the implement of specialized hardware. The geometric principle focuses on the geometric correction of curved images. This type of method usually design various geometric algorithms, such as Curvilinear Projection \cite{meng2015extraction,meng2018exploiting}, Covariance Matrix Adaptation \cite{moravec2016automatic}, Low-rank Textures \cite{yue2019distorted}, etc. These algorithms usually target specific distortion shapes, such as the general cylindrical distortion \cite{meng2018exploiting}, the barrel distortion \cite{moravec2016automatic}, the grouped circular arcs distortion \cite{zhang2019line}, etc. The above two types of methods aims at the regular deformation in natural scenes. However, different from the traditional task, the liquify of image editing is an irregular deformation operation, and there are various types of distortion in the image processing. Therefore, the restoration need to tackle the complexity and irregularity problems of the distortion. This requires seeking an appropriate mapping for any distorted image and generating visually realistic results.

Image generation has achieved great success since the advances of GANs \cite{goodfellow2014generative}. This work has been applied in numerous fields like image inpainting \cite{yeh2017semantic}, image completion \cite{li2017generative}, image reconstruction\cite{zhang2019ranksrgan} and image restoration \cite{pan2020physics}. Recently GANs have made great progress in respect of perceived quality which can be divided into two categories. The first method focuses on the unsupervised setting, which employs the unpaired data from different domains to establish the cross-domain mapping \cite{zhu2017unpaired,shu2019co,yi2017dualgan}. The second method follows the supervised setting, which seeks the reconstruction mapping between the input image and target image in a pixel by pixel manner \cite{isola2017image,wang2018high,wang2018perceptual}. In this work, our purpose is to learn the mapping from the distorted image to the undistorted image, which requires paired images to train the model to produce the visually photo-realistic images.

GAN-based image restoration is currently applied in two tasks. The first task is to fill the masked or missed regions in images. This task usually uses the joint training by local and global losses to maintain the content coherence \cite{yang2017high,yu2018generative,li2017generative}. However, our task is a global consistency problem, so local generative and discriminative learning is useless. The second task is to remove the noise of the blurred image and reconstruct a high resolution result from a low resolution image. The common idea is to achieve super-resolution by pixel-to-pixel manner \cite{kupyn2018deblurgan,kupyn2019deblurgan,li2018learning}. But it is hard to restore the image only relying on the super-resolution model in our task. The detailed demonstration is in our experiment section. Our task is to seek the appropriate mapping that enables the image to be restored. This not only requires the solution to maintain the overall coherence of the image, but also requires that the generated image is perceptually convincing compared with the original one. Motivated by GAN's ability in inpainting and reconstruction, we implement the characteristics of GAN to achieve image restoration, especially for distorted image caused by liquify filter. Our work presents an effective baseline method by generative and discriminative learning, where a rectification network is adopted for correction and a refinement network is designed for completion.
\section{Dataset}
Distortion is an optical phenomenon for images through lens or in the distorted mirror. Unlike plane mirror, a distorting mirror has some non-planar regions, in which the mirror reflects a distorted image. Many image editing softwares are based on this optical principle to achieve the liquefy function, which produces pictures with special effects. However, since distortion causes aberration to image details and reduction of the image equality, backward image restoration is time-consuming and difficult to achieve.

In this paper, our proposed method is based on deep convolutional neural network mode. The deep convolutional neural network could extract more features as the layer increases and solve the complexed distortion problem. The model requires large scale dataset as support, consisting of training part and test part. The dataset should contain images of different distortion types and images of same distortion type, different distortion rates. It also need to have ground truth to guide the optimization. In this task, the most common types of distortion we come across are barrel (negative) distortion and pin cushion (positive) distortion. 

Since there is no public dataset suitable for this task, we need to create a dataset as a benchmark in our evaluation. Obtaining the required images in the real world is difficult. Additional image editing software is also able to achieve our goal. However, we need a large amount of images for training, and processing the images one by one consumes too much time. Thus, we consider to utilize image geometric transformation to batch generate distorted image dataset, where allows different images.

We choose the celebA dataset \cite{liu2015faceattributes} as our basic dataset for modification. This dataset contains a large amount of aligned and cropped face data, which is suitable as the distortion processing object. Imaging distortion is an image interpolation. We assume the coordinate of a pixel $m$ in the source image $I$ is $\left(u,v\right)$, and coordinate of pixel $m^{\prime}$ in the destination image $I^{\prime}$ is $\left(x,y\right)$ [2]. In the paper, we implement distortion in a polar coordinate. Corresponding polar coordinate of pixel $m$ and $m^{\prime}$ are $\left(r,\theta\right)$ and$ \left(r^{\prime},\theta^{\prime}\right)$, so we can obtain that radius from the center $r=\sqrt{x^2+y^2} $ and $\theta=\tan \frac{y}{x}$. We define the distortion as a forward mapping function from source pixel to destination pixel.
\begin{equation}
\left\{
\begin{aligned}
& r^\prime=k\cdot r,\ {\ \theta}^\prime=\ \theta \\
& x^{\prime}=r^\prime\cdot cos\theta^\prime \\
& y^{\prime}=\ r^\prime\cdot sin\theta^\prime
\end{aligned}
\right.
\end{equation}

\begin{figure}[h]
	\centering
	\subfigure[]{
		\includegraphics[width=0.14\textwidth]{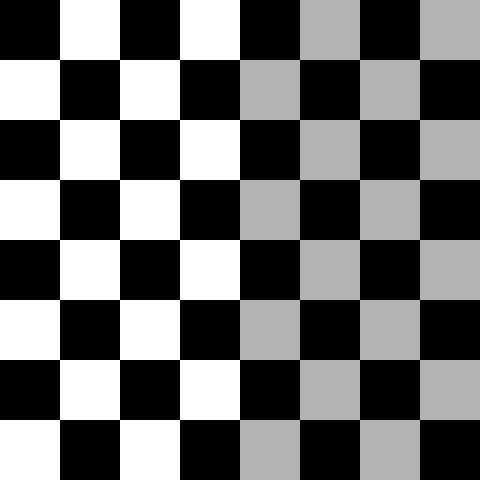}
	}
	\subfigure[]{
		\includegraphics[width=0.14\textwidth]{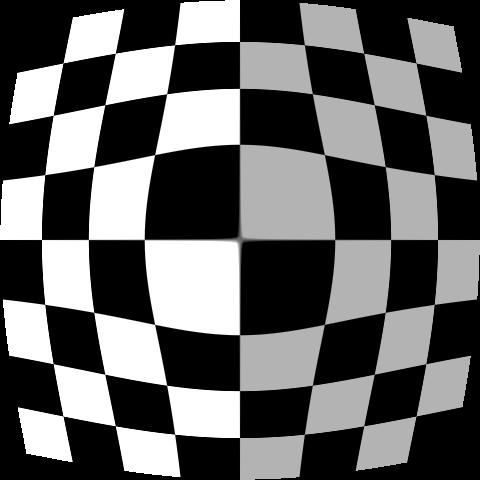}
	}
	\subfigure[]{
		\includegraphics[width=0.14\textwidth]{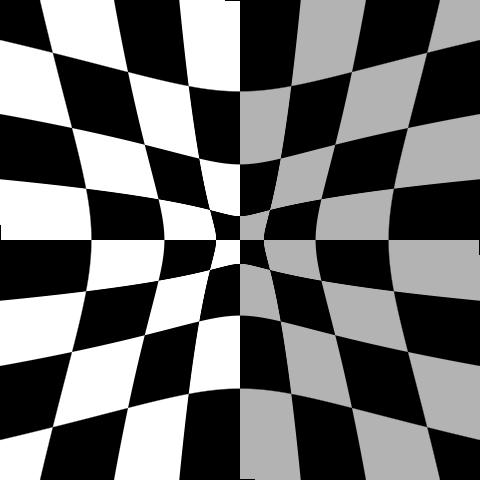}
	}
	\caption{Image distortion result. From left to right:(a) original image. (b) negative distortion. (c) positive distortion.}	
	\label{distortionresult}
\end{figure}

We call parameter $k$ the scaling factor for parametric mappings. When $k=1$,  the processed image stays the same as the original one. When $k\neq1$, the function distorts the image. If $0<k<1$, it performs convex distortion. The selected region expands, that image magnification decreases with distance from the center. Pixels within the selected region takes the value from pixels, whose coordinate has the same angel and larger radius. If $k>1$, it performs concave mirror imaging. The selected region shrinks, that image magnification increases with the distance from the center. Pixels within the selected region take the values from pixels, whose coordinate have the same angel and smaller radius. After distortion, we convert the polar coordinate into Cartesian coordinate and get the distorted image. Figure \ref{distortionresult} illustrates our distortion result.

The proposed optical distortion image dataset, Distorted Face Dataset (DFD), consists of 202,599 images: 100,000 negative distortion ones and 102,599 positive distortion ones. Parameter $k$ for negative and positive distortion take value 0.5, 0.8 and 1.5, 2.7 separately. Each image is 218$\times$178$\times$3 pixels. We should notice that since the distorted image itself is unnatural, our data does not need to be augmented to reduce the gap between real and synthetic data. This new dataset will be employed as a benchmark to evaluate the restoration model for the distorted image restoration task.

\section{Method}
Figure \ref{framework} illustrates the overview of our restoration framework. Our method consists of two steps: rectification and refinement. We first get the corrected prediction through the rectification network, then the rough prediction is fed into the refinement network to generate high frequency result. Each stage adopts a discriminator to obtain the precise reconstruction mapping. We demonstrate each component in the following.

\begin{figure}
	\centering
	\includegraphics[scale=0.5]{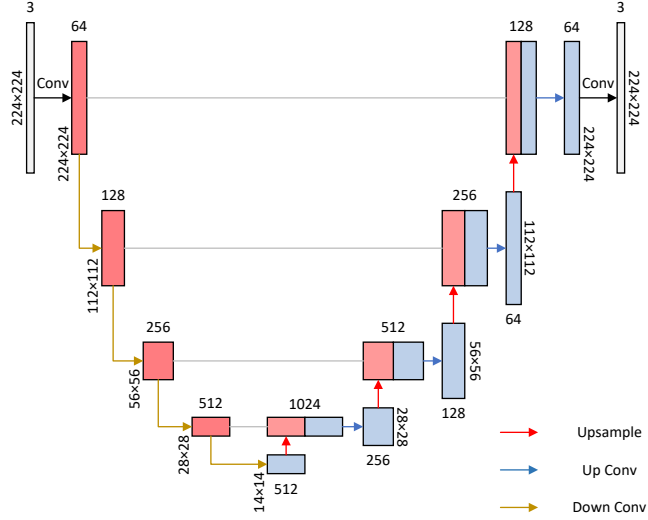}
	\caption{Illustration of rectification network. Four downsampling convolution layers are first performed to yield the feature map at 1/16 scale, then the four upsampling convolution layers restore the result to the same scale as the input.}
	\label{rectification}
\end{figure}

\subsection{Rectification Network}
The purpose of rectification network is to learn a non-linear transformation from warped space to normal space, which requires pixel-wise supervision. We follow the architectural guidelines of U-Net \cite{ronneberger2015u} which proves the effectiveness and simplicity in the pixel-wise prediction of the semantic segmentation, as shown in Figure \ref{rectification}. U-Net is based on the fully convolutional network \cite{long2015fully} with a series of downsampling steps and upsampling steps. The upsampling feature maps are concatenated with the downsampling feature maps. The difference is that our feature map is down sampled to 512 channels due to the computational overhead. Each upsampling step consists of an upsampling layer that double the size of feature map by bilinear interpolation and two repeated 3$\times$3 convolution layers (BN) that yield the number of feature channel at 1/4 scale. Each convolution layer is followed by a batch normalization layer and a rectified linear unit (ReLU). We also modify the padding scheme of each convolution layer to keep the input image and output image the same size.

\subsection{Refinement Network}
Only a simple rectification network may not generate satisfactory results. Since cascaded refinement has verified the effectiveness of image generation \cite{chen2017photographic}, we design a refinement network to produce missing details and preserve overall consistency, as shown in Figure \ref{refinement}. Motivated by the framework proposed by Johnson et al. \cite{johnson2016perceptual}, the refinement network is made up of two 7$\times$7 convolution layers, a 3$\times$3 convolution layer, two downsample blocks, five residual blocks \cite{he2016deep} and two upsample blocks. Each downsample block doubles the number of feature channel while yielding the feature map at 1/2 scale, which is composed of a 3$\times$3 convolution layer, a BN layer and PReLU activation. The residual block contains two 3$\times$3 convolution layers with 256 channels, two BN layers and PReLU activation. Each upsample block consists of a 3$\times$3 deconvolution layer halving the feature channel number, a BN layer and PReLU activation. Besides, to make the model generalize better and train faster, we adopt the local skip connection in the residual block and the global skip connection between input and output.

\subsection{Discrimination Network}
To ensure that the prediction is visually realistic and coherent, we utilize the discrimination network to discriminate generated images and original images. Distorted image restoration is a global consistency task, which requires the generated content to be consistent with the surrounding contexts. Therefore, we adopt the global discriminator where the architecture is the same as \cite{ledig2017photo} to keep the whole image perceptually consistent, as shown in Figure \ref{discriminator}. It applies eight convolutional layers that yield the feature map at 1/8 scale and 512 output channels as in the VGG network \cite{simonyan2014very}. It obtains the final probability through two dense layers and a sigmoid activation. We employ the global discriminator in the rectification generator and the refinement generator to improve the warpage and coherence, respectively.

\subsection{Loss}
In each step, training is an adversarial process introduced by \cite{goodfellow2014generative}, where the generator $G$ competes with the discriminator $D$ and the two networks are optimized alternately. The generator $G$ aims to generate content that can deceive the discriminator $D$ while the goal of the discriminator $D$ is to distinguish between generated samples and real samples. The game between them can be defined as a minimax objective function:

\begin{figure}
	\centering
	\includegraphics[scale=0.6]{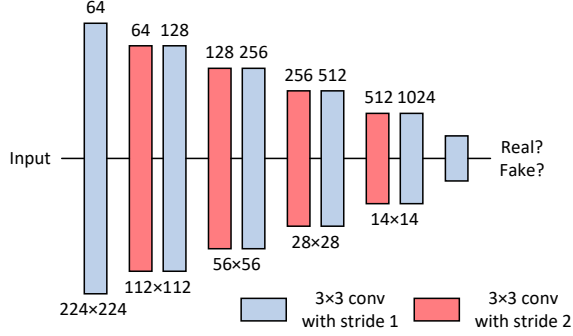}
	\caption{Architecture of discriminator network. The number below the convolution layer represents the size of feature map and the number above represents the channel.}
	\label{discriminator}
\end{figure}

\begin{equation}
\max \limits_{\theta_{D}} \min \limits_{\theta_{G}} \mathbb{E}_{\tilde{z} \sim \mathbb{P}_{d}}[\log(D_{\theta_{D}}(\tilde{z}))]+\mathbb{E}_{z \sim \mathbb{P}_{g}}[1-\log(D_{\theta_{D}}(G_{\theta_{G}}(z)))].
\label{gan}
\end{equation}where $\theta_{D}$ denotes the parameter of the discriminator $D$, $\theta_{G}$ denotes the parameter of the generator $G$, $\tilde{z}$ is the target sample subjecting to the real data distribution $\mathbb{P}_{d}$ while $z$ is the distorted sample subjecting to the simulated data distribution $\mathbb{P}_{g}$. We adopt the content and adversarial loss from some GAN approaches \cite{bai2018finding} \cite{isola2017image} to optimize the generator network. 

Content loss. A classic choice for measuring the perceptual similarity between restored images and original images is the pixel-wise L2 or MSE loss. This loss enforces the output image produced by the generator to be close to the ground truth, further bridging the input-output gap, formulated as:
\begin{equation}
\mathcal{L}_{mse}=\frac{1}{CWH}\sum_{i=1}^{N}\left \|\tilde{z}^{(i)}-G_{\theta_{G}}(z^{(i)})  \right \|_{2}^{2},
\label{Pixel-wiseloss}
\end{equation}where $G_{\theta_{G}}(z^{(i)})$ is the generated image and $C$,$W$,$H$ correspond to the channel, weigh and height of the image.

Adversarial loss. Since the goal of the discriminator is to distinguish generated images from real images, the adversarial loss is adopted to encourage the generator to produce the natural-looking images that manage to fool the discriminator, formulated as:
\begin{equation}
\mathcal{L}_{adv}=\sum_{i=1}^{N}- \log D_{\theta_{D}}(G_{\theta_{G}}(z^{(i)})),
\label{Adversarialloss}
\end{equation}where $D_{\theta_{D}}(G_{\theta_{G}}(z^{(i)})$ represents the probability of the discriminator over the prediction.

Based on above analysis, the content loss is utilized to restore general content while the adversarial loss focuses on completing texture details. The overall loss function can be formulated as:
\begin{equation}
\mathcal{L}=\mathcal{L}_{mse}+\lambda \mathcal{L}_{adv}.
\end{equation}where $\lambda $ is the trade-off weight, which is set by the cross-validation experiments. The discriminator of each step shares the same formulation of the loss function.
\begin{figure*}[htbp]
	\centering
	\subfigure{
		\includegraphics[width=0.1\textwidth]{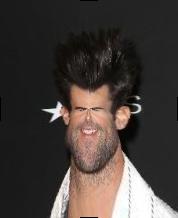}
		\includegraphics[width=0.1\textwidth]{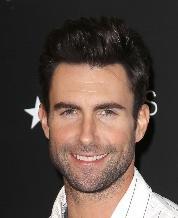}
		\includegraphics[width=0.1\textwidth]{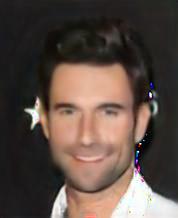}
		\includegraphics[width=0.1\textwidth]{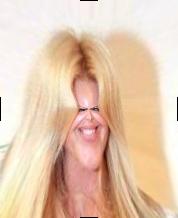}
		\includegraphics[width=0.1\textwidth]{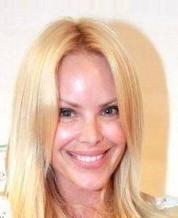}
		\includegraphics[width=0.1\textwidth]{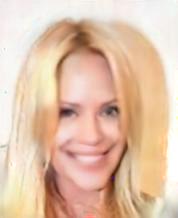}
		\includegraphics[width=0.1\textwidth]{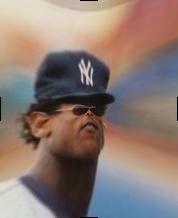}
		\includegraphics[width=0.1\textwidth]{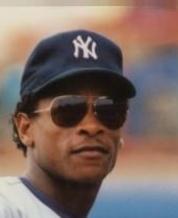}
		\includegraphics[width=0.1\textwidth]{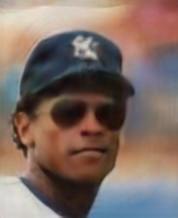}
	}
	\subfigure{
		\includegraphics[width=0.1\textwidth]{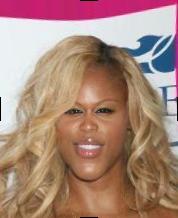}
		\includegraphics[width=0.1\textwidth]{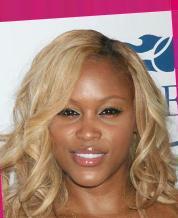}
		\includegraphics[width=0.1\textwidth]{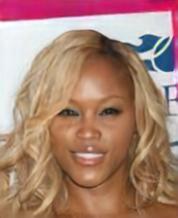}
		\includegraphics[width=0.1\textwidth]{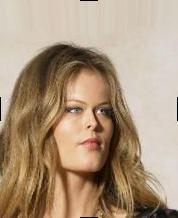}
		\includegraphics[width=0.1\textwidth]{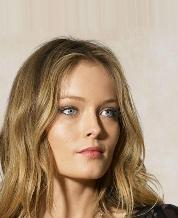}
		\includegraphics[width=0.1\textwidth]{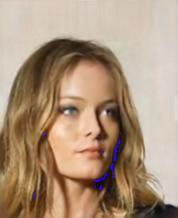}
		\includegraphics[width=0.1\textwidth]{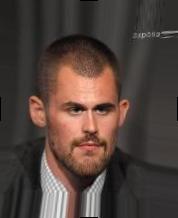}
		\includegraphics[width=0.1\textwidth]{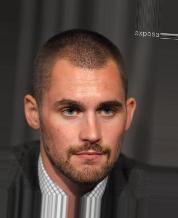}
		\includegraphics[width=0.1\textwidth]{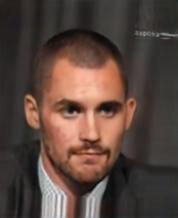}
	}
	\subfigure{
		\includegraphics[width=0.1\textwidth]{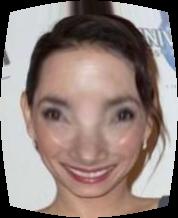}
		\includegraphics[width=0.1\textwidth]{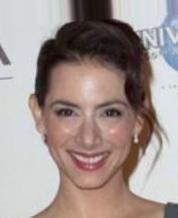}
		\includegraphics[width=0.1\textwidth]{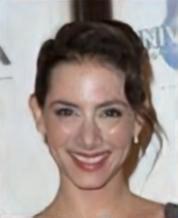}
		\includegraphics[width=0.1\textwidth]{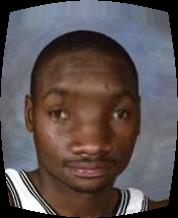}
		\includegraphics[width=0.1\textwidth]{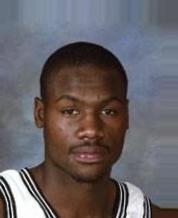}
		\includegraphics[width=0.1\textwidth]{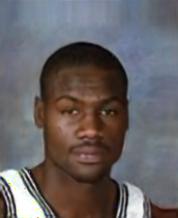}
		\includegraphics[width=0.1\textwidth]{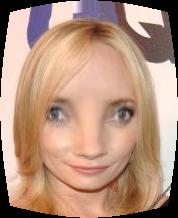}
		\includegraphics[width=0.1\textwidth]{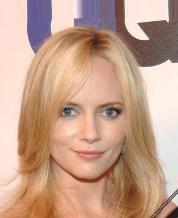}
		\includegraphics[width=0.1\textwidth]{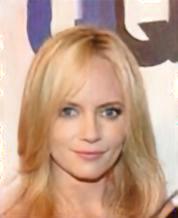}
	}
	\subfigure{
		\includegraphics[width=0.1\textwidth]{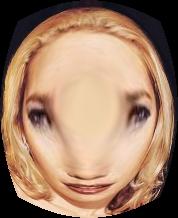}
		\includegraphics[width=0.1\textwidth]{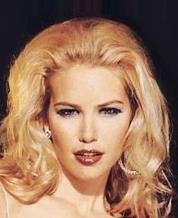}
		\includegraphics[width=0.1\textwidth]{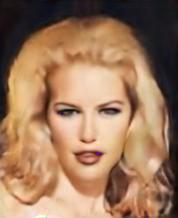}
		\includegraphics[width=0.1\textwidth]{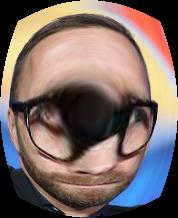}
		\includegraphics[width=0.1\textwidth]{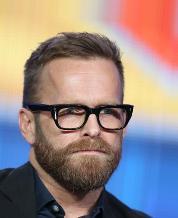}
		\includegraphics[width=0.1\textwidth]{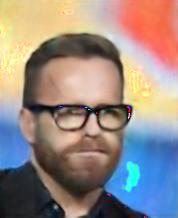}
		\includegraphics[width=0.1\textwidth]{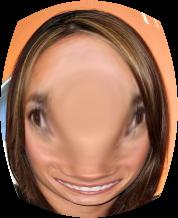}
		\includegraphics[width=0.1\textwidth]{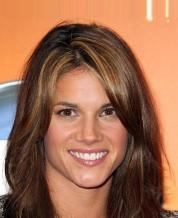}
		\includegraphics[width=0.1\textwidth]{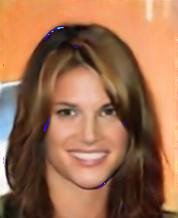}
	}
	\caption{Distorted face restoration results on the DFD benchmark. In each panel from left to right: distorted image, original image and our method result.}	
	\label{result}
\end{figure*}
\section{Experiment}
The purpose of our extensive experiments is to demonstrate the performance of our method as a simple baseline. We expect to demonstrate both effectiveness and limitation of our method and analyze the contribution of each module. We evaluate our method on the proposed benchmark and the application.

\begin{table}
	\centering
	\caption{Quantitative evaluations in terms of PSNR at four different categories of distortion S1-S4 and the overall benchmark. Higher values are better.}
	\label{table1}
	\begin{tabular}{|l|c|c|c|c|c|}
		\hline
		& S0 & S1 & S2 & S3 & S4 \\ \hline
		Our method & \textbf{25.12} & \textbf{23.57} & \textbf{26.42} & \textbf{25.99} & \textbf{24.49} \\ \hline
		Only Rectification & 21.42 & 19.43 & 24.33 & 22.37 & 19.54 \\ \hline
		Only Refinement & 17.36 & 16.82 & 17.99 & 17.62 & 17.01 \\ \hline
		Refinement w/o $\mathcal{L}_{adv}$ & 17.17 & 16.96 & 17.78 & 17.41 & 16.52 \\ \hline
		Refinement w/o $\mathcal{L}_{mse}$ & 5.49 & - & - & - & - \\ \hline
		Rectification w/o $\mathcal{L}_{adv}$ & 18.93 & 18.01 & 19.77 & 19.27 & 18.68 \\ \hline
		Rectification w/o $\mathcal{L}_{mse}$ & 6.83 & - & - & - & - \\ \hline
	\end{tabular}
\end{table}

\begin{table}
	\centering
	\caption{Quantitative evaluations in terms of SSIM at four different categories of distortion S1-S4 and the overall benchmark. Higher values are better.}
	\label{table2}
	\begin{tabular}{|l|c|c|c|c|c|}
		\hline
		& S0 & S1 & S2 & S3 & S4 \\ \hline
		Our method & \textbf{0.813} & \textbf{0.782} & \textbf{0.856} & \textbf{0.834} & \textbf{0.779} \\ \hline
		Only Rectification & 0.773 & 0.731 & 0.839 & 0.802 & 0.721 \\ \hline
		Only Refinement & 0.544 & 0.542 & 0.563 & 0.544 & 0.528 \\ \hline
		Refinement w/o $\mathcal{L}_{adv}$ & 0.730 & 0.706 & 0.764 & 0.751 & 0.701 \\ \hline
		Refinement w/o $\mathcal{L}_{mse}$ & 0.072 & - & - & - & - \\ \hline
		Rectification w/o $\mathcal{L}_{adv}$ & 0.742 & 0.711 & 0.789 & 0.763 & 0.706 \\ \hline
		Rectification w/o $\mathcal{L}_{mse}$ & 0.061 & - & - & - & - \\ \hline
	\end{tabular}
\end{table}

\subsection{Implementation Details} 
The benchmark contains both the original images and the distorted ones. Since the dataset is huge enough for deep learning, we split the DFD into training set and test set with 196K images for training and 4K images for testing. We ensure that no image exists in the test set and training set simultaneously in order to better evaluate the generalization of our method. All images for training and testing are of the size 224$\times$224. We start pre-training strategy to train for 10 epochs, initializing the trade-off weight $\lambda$ as 0.001 and initializing the learning rate as $0.01$. All experiments are conducted on 8 NVIDIA RTX 2080Ti GPUs, CUDA 10.1 and cuDNN 7.6.5.

We explore the restoration performance through both qualitative and quantitative analysises. The qualitative evaluation mainly relies on visual perception while the quantitative evaluation utilizes two metrics to quantify the performance. The first metrics is peak signal to noise ratio (PSNR), which is based on the difference between corresponding pixel values. Since PSNR is based on error-sensitive image quality evaluation, it does not consider the visual characteristics of the human eye. There are often cases that the evaluation result is inconsistent with the subjective feeling of the person. So we introduce the second metrics called structural similarity (SSIM), which measures the holistic similarity from three aspects: brightness, contrast, and structure.
\begin{figure}
	\centering
	\subfigure{
		\begin{minipage}[b]{0.1\textwidth}
			\includegraphics[width=1\textwidth]{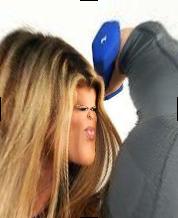} \vspace{4pt}
			\includegraphics[width=1\textwidth]{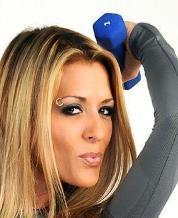} \vspace{4pt}
			\includegraphics[width=1\textwidth]{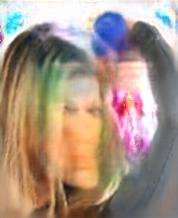} \vspace{4pt}
			\includegraphics[width=1\textwidth]{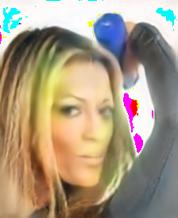} \vspace{4pt}
			\includegraphics[width=1\textwidth]{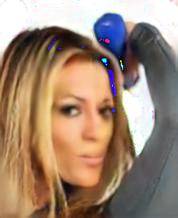}
		\end{minipage}
	}
	\subfigure{
		\begin{minipage}[b]{0.1\textwidth}
			\includegraphics[width=1\textwidth]{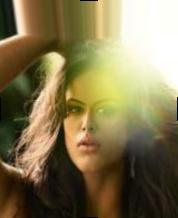} \vspace{4pt}
			\includegraphics[width=1\textwidth]{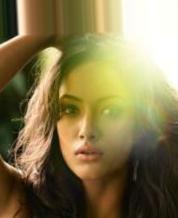} \vspace{4pt}
			\includegraphics[width=1\textwidth]{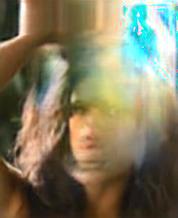} \vspace{4pt}
			\includegraphics[width=1\textwidth]{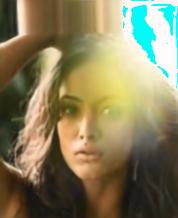} \vspace{4pt}
			\includegraphics[width=1\textwidth]{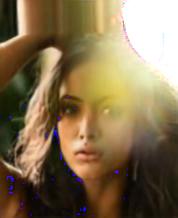} 
		\end{minipage}
	}
	\subfigure{
		\begin{minipage}[b]{0.1\textwidth}
			\includegraphics[width=1\textwidth]{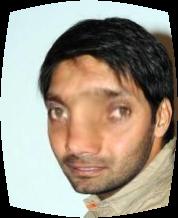} \vspace{4pt}
			\includegraphics[width=1\textwidth]{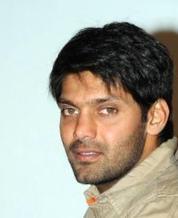} \vspace{4pt}
			\includegraphics[width=1\textwidth]{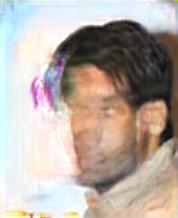} \vspace{4pt}
			\includegraphics[width=1\textwidth]{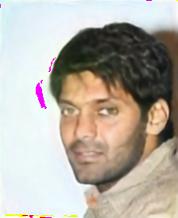} \vspace{4pt}
			\includegraphics[width=1\textwidth]{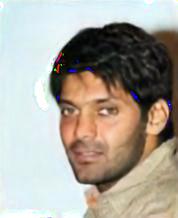} 
		\end{minipage}
	}
	\subfigure{
		\begin{minipage}[b]{0.1\textwidth}
			\includegraphics[width=1\textwidth]{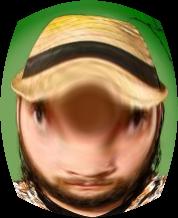} \vspace{4pt}
			\includegraphics[width=1\textwidth]{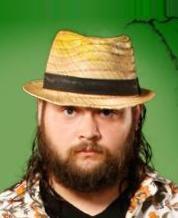} \vspace{4pt}
			\includegraphics[width=1\textwidth]{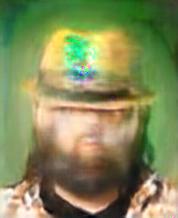} \vspace{4pt}
			\includegraphics[width=1\textwidth]{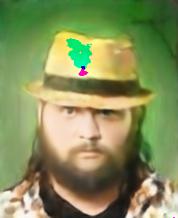} \vspace{4pt}
			\includegraphics[width=1\textwidth]{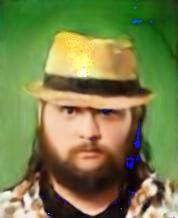} 
		\end{minipage}
	}
	\caption{Visual comparisons of DFD restoration result. In each column from top to bottom: distorted image, original image, only refinement result, only rectification result and our method result.}	
	\label{comparision}
\end{figure}

\subsection{Qualitative Results} 
Figure \ref{result} shows our restoration results on the DFD. The results in the third column of each panel in the figure show that our model obtains the perceptually convincing images and has good generalization ability on a variety of distorted images. We also compare our model with the single rectification network and the single refinement network, the results in Figure \ref{comparision} show that our method performs better than other methods and achieves the most visually satisfactory results. It is worth noting that the deeper the image distortion degree is, the more details are lost in the restored image. For example, the restoration of the eye part of some images is not clear enough. The reason is that the eye part is located at the center of the distortion, which is the most severely distorted. Overall, our method can correct various distortions of the image and partially or completely supplement the missing details.

\subsection{Quantitative Results} 
In order to further explore the relationship between the distortion and the model, we report our quantitative evaluation on the benchmark in terms of PSNR and SSIM, as shown in Tables \ref{table1}-\ref{table2}. We separately calculate two metrics under the overall test set S0 and each distortion category where S1-S4 correspond to four distortion parameters at 0.5, 0.8, 1.5 and 2.7 respectively. The results show that our model illustrates good generalization on whether positive or negative distorted images. However, as the degree of distortion increases, the restoration performance of our model decreases in both metrics. We then compare our model with the single rectification network and the single refinement network. We can conclude that our method performs better than others and achieves the highest numerical performance. This is due to the nature of our method, which can seek a suitable nonlinear mapping for reconstruction and the corresponding supplement to correct the image.
\begin{figure}
	\centering
	\subfigure{
		\begin{minipage}[b]{0.1\textwidth}
			\includegraphics[width=1\textwidth]{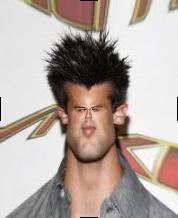} \vspace{4pt}
			\includegraphics[width=1\textwidth]{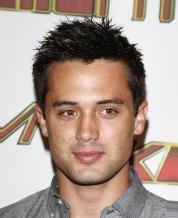} \vspace{4pt}
			\includegraphics[width=1\textwidth]{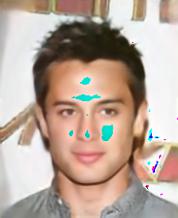} \vspace{4pt}
			\includegraphics[width=1\textwidth]{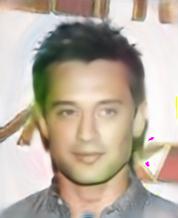} \vspace{4pt}
			\includegraphics[width=1\textwidth]{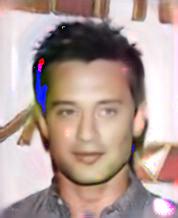}
			\includegraphics[width=1\textwidth]{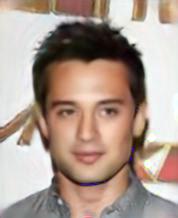}
		\end{minipage}
	}
	\subfigure{
		\begin{minipage}[b]{0.1\textwidth}
			\includegraphics[width=1\textwidth]{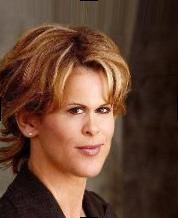} \vspace{4pt}
			\includegraphics[width=1\textwidth]{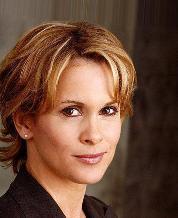} \vspace{4pt}
			\includegraphics[width=1\textwidth]{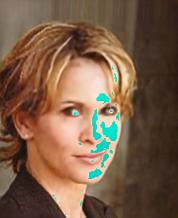} \vspace{4pt}
			\includegraphics[width=1\textwidth]{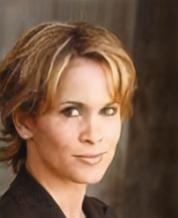} \vspace{4pt}
			\includegraphics[width=1\textwidth]{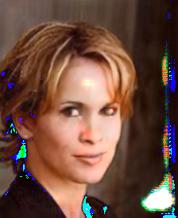}
			\includegraphics[width=1\textwidth]{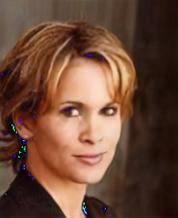}
		\end{minipage}
	}	
	\subfigure{
		\begin{minipage}[b]{0.1\textwidth}
			\includegraphics[width=1\textwidth]{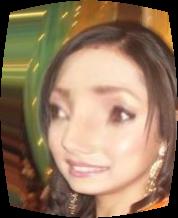} \vspace{4pt}
			\includegraphics[width=1\textwidth]{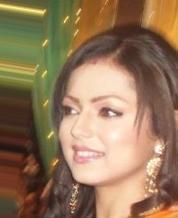} \vspace{4pt}
			\includegraphics[width=1\textwidth]{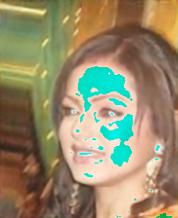} \vspace{4pt}
			\includegraphics[width=1\textwidth]{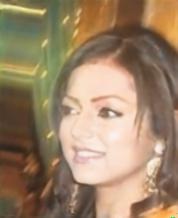} \vspace{4pt}
			\includegraphics[width=1\textwidth]{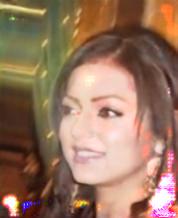}
			\includegraphics[width=1\textwidth]{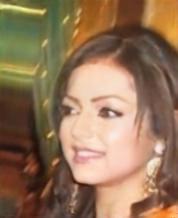}
		\end{minipage}
	}	
	\subfigure{
		\begin{minipage}[b]{0.1\textwidth}
			\includegraphics[width=1\textwidth]{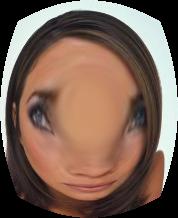} \vspace{4pt}
			\includegraphics[width=1\textwidth]{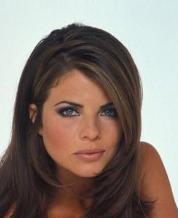} \vspace{4pt}
			\includegraphics[width=1\textwidth]{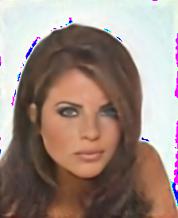} \vspace{4pt}
			\includegraphics[width=1\textwidth]{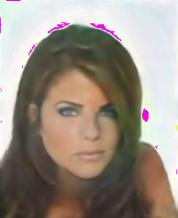} \vspace{4pt}
			\includegraphics[width=1\textwidth]{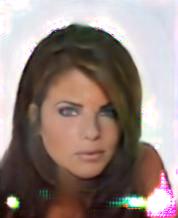}
			\includegraphics[width=1\textwidth]{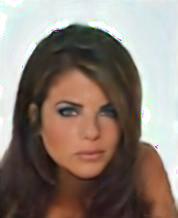}
		\end{minipage}
	}
	\caption{Ablation studies on DFD. In each column from top to bottom: distorted image, original image, only rectification result without adversarial loss, only rectification result, our method result without adversarial loss and our method result.}	
	\label{ablation}
\end{figure}

\subsection{Effect of Refinement Network} 
We drop out the refinement network and purely rely on the rectification network to produce the results. Figure \ref{comparision} presents the generated images and Tables \ref{table1}-\ref{table2} further shows the quantitative comparisons. It demonstrates that the cascaded refinement can further optimize the perceptual quality. It denoises the prediction and produce high frequency details in color and contrast. By implementing more content completion, the results are more coherent with the target compared with training without the rectification network. 

\subsection{Effect of Rectification Network} 
We omit the rectification network and only apply the refinement network in the optimization. In Tables \ref{table1}-\ref{table2}, only Refinement Network training acquires the lowest numerical performance in PSNR and SSIM. And the third row in Figure \ref{result} demonstrates that only super-resolution model cannot match our task, mapping reconstruction is also essential in the restoration work. We can conclude that utilizing more reconstruction term can result in partly restoration for distorted image but does not guarantee the overall image content is correct.

\subsection{Effect of Adversarial Loss} 
We investigate the effectiveness of the adversarial loss in the two stages. To make the comparisons, we train the model ignoring the adversarial loss in the refinement network and rectification network, respectively. In Tables \ref{table1}-\ref{table2} the two metrics are lower than those of training with the adversarial loss. Also from Figure \ref{ablation}, we can see that without the adversarial loss, our method is still able to predict the restored image, but the results have noise and lose some details. The adversarial loss serves as an inpainting term for restoration, which is beneficial to capture the structure and encourages the network to produce sharp content.

\subsection{Effect of Content Loss} 
As shown in Table \ref{table1}-\ref{table2}, without the content loss to guide the optimization, the result of restoration is completely incorrect, neither in the rectification network nor in the refinement network. It means that the content loss is essential to the quality of the restoration result. The initial prediction without content loss is blurrier comparing with training with the content loss, which leads to the failure of the final prediction.

\subsection{Application} 
\begin{figure}[h]
	\centering
	\subfigure[]{
		\includegraphics[width=0.1\textwidth]{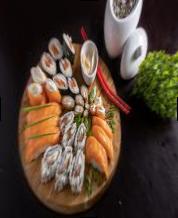}
		\includegraphics[width=0.1\textwidth]{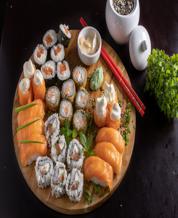}
		\includegraphics[width=0.1\textwidth]{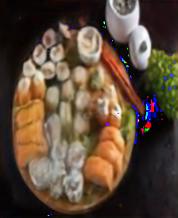}
	}
	\subfigure[]{
		\includegraphics[width=0.1\textwidth]{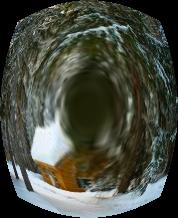}
		\includegraphics[width=0.1\textwidth]{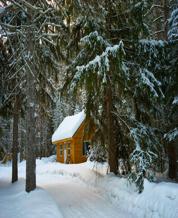}
		\includegraphics[width=0.1\textwidth]{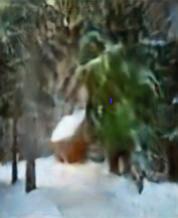}
	}
	\subfigure[]{
		\includegraphics[width=0.1\textwidth]{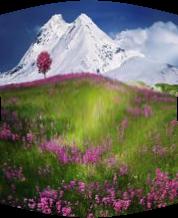}
		\includegraphics[width=0.1\textwidth]{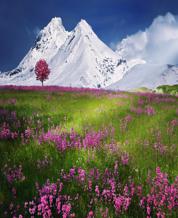}
		\includegraphics[width=0.1\textwidth]{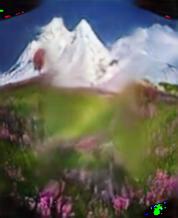}
	}
	\caption{Application results. From left to right: distorted image, original image and restoration image. From top to bottom: (a) concave effect, (b) convex effect, (c) a mixture of convex and concave effect.}	
	\label{application}
\end{figure}

Since the baseline method effectively seeks the appropriate warping and supplements the missing details, it can also be applied on other types of images besides face images. Here, we find some pictures from the Pexel website and demonstrate the versatility of our method from two aspects: 1) We utilize the distorted mirror effect to generate arbitrary rate of distortion. 2) We apply both convex and concave effects on the same picture. Figure \ref{application} shows several interesting restoration examples. Since the deep learning model is data-driven and our dataset only has concave and convex distortions, our model cannot generalize on other special effects such as sinusoid, affine, projective and piecewise linear. But the generalization of our method we have verified demonstrates that if sufficient training data is provided, the model is able to generalize on other types of distortions.

\begin{figure}[h]
	\centering
	\subfigure{
		\includegraphics[width=0.1\textwidth]{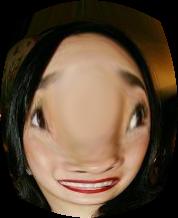}
		\includegraphics[width=0.1\textwidth]{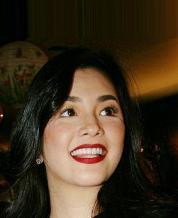}
		\includegraphics[width=0.1\textwidth]{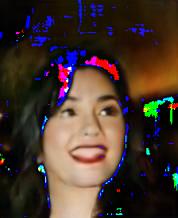}
	}
	\subfigure{
		\includegraphics[width=0.1\textwidth]{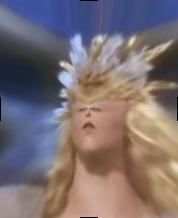}
		\includegraphics[width=0.1\textwidth]{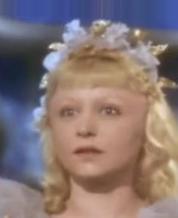}
		\includegraphics[width=0.1\textwidth]{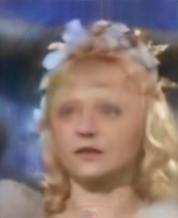}
	}
	\caption{The limitations of our method. From left to right: distorted image, original image and restoration image.}	
	\label{limitation}
\end{figure}

\subsection{Limitation} 
Although our method can achieve semantically plausible and visually satisfactory results as a baseline, some limitations still exist. When the distortion degree is too deep, the performance of restoration is not satisfactory. It is difficult for the algorithm to seek a suitable mapping. Some components of the face are not fully restored, and one unsatisfactory result is shown in the top row of figure \ref{limitation}. Besides, some images are not smooth enough after restoration, and noise is generated at the edge of the face, as shown in the second row of Figure \ref{limitation}. The reason is that in areas where the image is severely distorted or object edges are blurred, our network extract few feature resulting in loss of details. But since results are still visually pleasant, we may probably solve such cases by user interaction with image editing software.

\section{Conclusion}
In this paper, we explore the new distorted image restoration task caused by liquify filter. In order to find out the expected and unexpected innovations, we release a Distorted Face Dataset (DFD) containing negative and positive distortions. Motivated by the significant progress of GANs, we propose the first deep neural network by generative and discriminative learning for this task. The method is divided into two stages, rectification and refinement. The rectification stage learns the mapping to correct the distorted image and the refinement stage further optimizes the perceptual quality. Extensive experiments on the benchmark and the application have demonstrated the effectiveness of our proposed method. However, certain bottlenecks still exist in our work which come from two areas. One area is that the existing method doesn't restore the full perceptual distortion and the result still lacks high-frequency details when the distortion degree is too deep. The other area is that the data-driven model is required to continuously learn the input feature so that the model can generalize and restore on unknown distortions, since the liquify filter can result in various distortions. In our future work we will focus on these areas leading to a continuously learning model producing more perceptually superior images. This method is a basic first step, but we expect some more innovative method will be inspired.

\section*{Acknowledgements}
This work is partially supported by the National Key R\&D Program of China (No.2018YFB0105203), NSFC (No.61932014, 61972246), Natural Science Foundation of Shanghai (No.18ZR1418500) and Research Collaboration Grant from NII, Japan.

\clearpage
{\small
	\bibliographystyle{unsrt}  
	\bibliography{mylib}
}
\end{document}